# Mamyshev oscillator based on gain-managed nonlinearity and chirped pulse amplification

**Baofu Zhang[1], Sihua Lu[2], Qiurun He[3], Ya Lin[3], Zhongxing Jiao[3], and Biao Wang[1,4,5]**

[1] *Research Institute of Interdisciplinary Science & School of Materials Science and Engineering, Dongguan University of Technology, Dongguan 523808, China*

[2] *EMGO-TECH Ltd., Zhuhai 519000, China*

[3] *School of Electronics and Communication Engineering, Sun Yat-sen University, Guangzhou 510275, China*

[4] *School of Physics, Sun Yat-sen University, Guangzhou 510275, China*

[5] *Faculty of Mechanical Engineering & Mechanics, Ningbo University, Ningbo 315211, China*

**Abstract** We experimentally demonstrate a Mamyshev oscillator based on gain-managed nonlinearity and chirped pulse amplification. Different from other Mamyshev oscillators, the gain-managed nonlinear regime serves as a seed provider instead of a power amplifier in one arm of this laser. The output pulse energy over 300 nJ and pulse width of 739 fs has been achieved from the chirped pulsed amplification in another arm. This configuration provided a new approach to design a high-energy ultrafast laser.

## I. INTRODUCTION

---

Correspondence to: No. 1 Daxue Road, Songshanhu District, Dongguan, Guangdong Province, Chine. Email: zhangbf@dgut.edu.cn

Benefit from compact structure, excellent heat dissipation, low-cost designs and high-peak intensity, ultrafast fiber lasers have played important roles in industrial, scientific and biomedical applications [1-3]. To obtain high-energy output and avoid pulse distortion in these applications, most laser sources are constructed based on chirped-pulse amplification (CPA) [4]. In a typical CPA system, laser pulses from an oscillator are first stretched to reduce the peak intensity and hence large nonlinearities introduced by the fiber structure. The stretched pulses are amplified, and then compressed to several hundred femtoseconds. In order to achieve higher peak power and shorter pulse duration for some applications such as high harmonic generation and microscopy [5-7], ultrafast fiber lasers based on Mamyshev oscillator (MO) and gain-managed nonlinear amplification (GMNA) have been demonstrated [8, 9]. In contrast to CPA systems, these two configurations take advantage of nonlinear effects, and they can produce sub-100-fs pulses beyond gain-narrowing limit.

Many efforts have been devoted to improving the output characteristics of MOs and GMNA systems [10-18]. However, most of these systems with commercial single-mode fiber can only produce sub-200-nJ level due to the strength of nonlinear effects especially stimulated Raman scattering (SRS) [9, 10, 14]. This is an inherent limitation in MOs, GMNA systems and other nonlinear amplifiers: high-peak-power pulses in a small core with long propagation distances results in the accumulation of large nonlinear effects. Although the output pulse energy of MOs and GMNA systems can be higher by using large-mode-area (LMA) fibers [11-13, 16-18], it is still much lower than the output pulse energy of CPA systems. Therefore, this limitation will also restrict the application of these two novel ultrafast lasers in industrial and medical fields that require higher pulse energy.

To overcome the inherent limitation in MOs and GMNA systems, more attention should be paid to their pulse evolution mechanisms. In fact, there is a close relationship between the MOs and GMNA systems since the gain-managed nonlinear (GMN) evolution can also be found in MOs [9, 19, 20]. In our previous work [20], we offer another perspective on understanding the MO in which different pulse evolutions are allowed in its two arms. One of its arms can be considered as a power amplifier seeded by a bandpass filtered laser, while another arm can be considered as the provider of seed laser. Therefore, we found that the GMN evolution beyond gain-narrowing limit in the power amplifier arm is the key to achieve high-peak-power output in MOs. However, most MOs are limited to this design approach, resulting in their low-pulse-energy output. On other hand, a nonlinear attractor underlies the GMNA regime [9]. Therefore, the GMN evolution is also suitable for the seed laser arm of MOs, while high pulse energy can be achieved in the power amplifier arm with the configuration similar to CPA systems.

Here, we present a MO based on gain-managed nonlinearity and CPA. To the best of our knowledge, this configuration is first demonstrated. Ultrashort pulses with their pulse energy over 300 nJ can be achieve from this laser.

## II. Methods

### Experimental setup

The experimental setup is a seed-injected ring-cavity MO with two active arms, and its schematic is depicted in Figure 1. Arm 1 is a typical GMNA configuration. The laser pulses from Arm 2 first pass through a fiberized bandpass filter (BF 1), and then they are coupled into a 3.5-m LMA ytterbium-doped fiber (GF 1, PLMA-YDF-10-125-M) via a combiner (Comb 1) with the multimode 976-nm pump laser (Pump 1). The central wavelength and the bandwidth (full width at half maximum, FWHM) of BF 1 are 1035 nm and 2.8 nm, respectively. Since there are self-starting problems of MOs [8], an ultrafast seed is coupled and used for starting the mode-locked operation of this MO. The seed can produce ultrashort pulses with 0.1-nJ pulse energy, 1-ps pulse duration and 3-nm bandwidth (FWHM) centered at 1030 nm. Its position is optimized and located between Arm 1 and Arm 2.

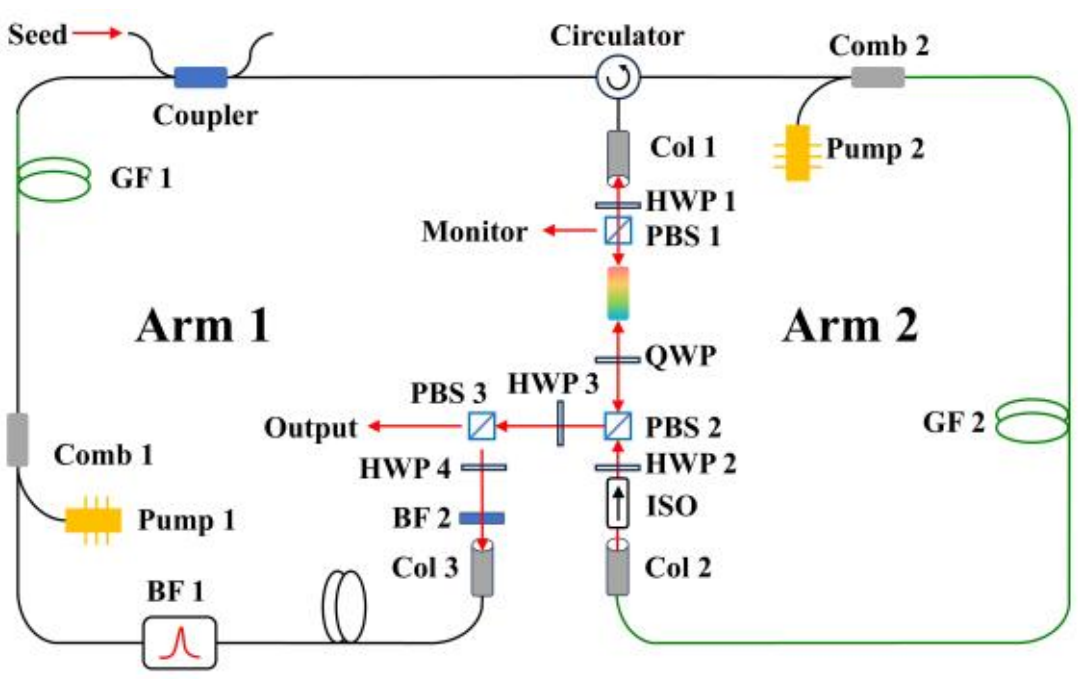

**Figure 1**. Experimental setup for the Mamyshev oscillator based on GMNA and CPA.

Arm 2 is a simplified and compact CPA configuration. The laser pulses from Arm 1 are launched into a chirped volume Bragg gratings (CVBG) via a fiber circulator, a collimator (Col 1), a half-wave plate (HWP 1) and a polarizing beam splitter (PBS 1). A small amount of laser output from the other side of PBS 1 is used to monitor the operation status of Arm 1. The 50-mm CVBG is linear chirp type with a stretching factor of 64 ps/nm. Its central wavelength is 1030 nm with a spectral bandwidth (FWHM) of 8 nm. Therefore, the CVBG serves not only as a stretcher but also as a bandpass filter for the input laser from Arm 1. After reflected by the CVBG, the chirped pulses pass through the CIR again, and then they are coupled into a 1.5-m LMA ytterbium-doped fiber (GF 2, PLMA-YDF-14-125-UF) via a combiner (Comb 2) with the multimode 976-nm pump laser (Pump 2). After power amplification in GF 2, the laser pulses are launched into the free space through a collimator (Col 2). In order to avoid nonlinear effects and ensure a large amplification factor in this CPA stage, GF 2 is not only the output fiber of Comb 2 but also the input fiber of Col 2. In the free space of Arm 2, an isolator is used to ensure unidirectional operation. The amplified pulses are coupled into another side of the CVBG which acts as a compressor after they pass through a half-wave plate (HWP 2), a polarizing beam splitter (PBS 2) and a quarter-wave plate (QWP). The high-peak-power dechirped pulses are output through a tunable coupler formed by the combination of a half-wave plate (HWP 3) and a polarizing beam splitter (PBS 3). The residual laser is launched into the ring cavity again via a half-wave plate (HWP 4), a bandpass filter (BF 2) and a collimator (Col 3). The central wavelength and the bandwidth (full width at half maximum, FWHM) of BF 2 are 1030 nm and 3.9 nm, respectively. The subsequent 6-m passive fiber are used to broaden laser bandwidth, and hence the laser pulses can pass through BF 1 in Arm 1 to achieve the operation of MO. All fibers used in the MO are polarization maintaining for the high environmental stability. The total cavity length is about 23.8 m, of which the fiber length is about 14.5 m and the free space length is about 2.6 m.

**Numerical simulation**

We performed numerical simulations of the MO by using the model presented in our previous work [20]. The second- and the third-dispersion, as well as Kerr nonlinearity, self-steeping effect and SRS are considered in the simulations. Longitudinal evolution profiles for the laser pulse in the MO and its spectra can be determined by iteratively solving the joint of generalized nonlinear Schrödinger equation and rate equations. Numerical results with accurate fiber parameters are shown in Figure 2 when the input power of Pump 1 and Pump 2 are set to be 1.9 W and 10 W, respectively. The pulse evolution in the gain fiber of two arms can be distinguished. In GF 1, the broadband spectrum beyond 1100 nm and the asymmetric pulse shape at the end of the gain fiber are presented, which are typical characteristics of GMN regime. In GF 2, the chirped pulse with narrow bandwidth has been power amplified to a high level while its pulse width and bandwidth remain unchanged.

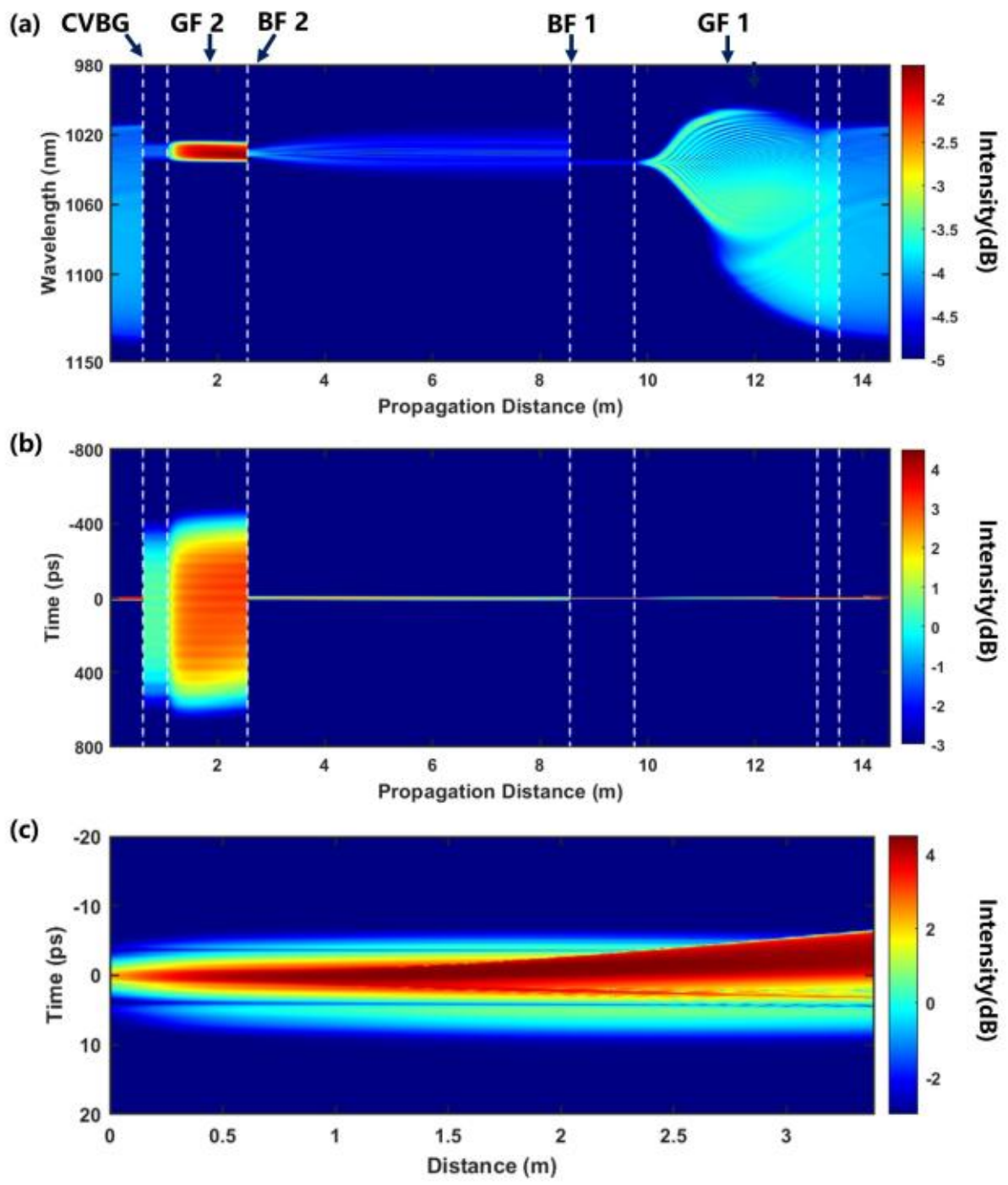

**Figure 2**. Simulated (a) spectral evolution and (b) temporal evolution of a laser pulse in the Mamyshev oscillator. (c) shows the detail of simulated temporal evolution in GF 1.

## III. Results

Experiments were performed with guidance from the simulations. The ultrafast seed was required at first to start the mode-locked operation of MO. In the following experiments, the output power of Pump 1 was kept constant at 1.9 W. When the output power of Pump 2 was increased to 3 W, unstable mode-locked pulses with the repetition rate of 12.6 MHz corresponding to the total cavity length could be found. Here, the seed was turned off, and then the stable single-pulse mode-locking operation could be achieved. As shown in Figure 3(a), the output power from PBS 3 of MO increased linearly with the increasing power of Pump 2. The slope efficiency is calculated to be 52.3%. Attribute to the nonlinear attractor of GMNA in Arm 1, the stable mode-locking operation could maintain when the power of Pump 2 was up to 10 W. The maximum output power of MO was measured to be 4.12 W, corresponding to the pulse energy of 327.0 nJ. Its output power stability with the root mean square of 0.3% could be found in Figure 3(b), indicating the long-term stable operation of MO. When the pump power reached a higher level, the MO exhibited a complex and unstable operation dynamic, and then mode-locking state would vanish. This instability at higher power could be introduced by stronger SRS effect in Arm 1 (shown in Figure 4(a)) or the gain-narrowing effect in Arm 2.

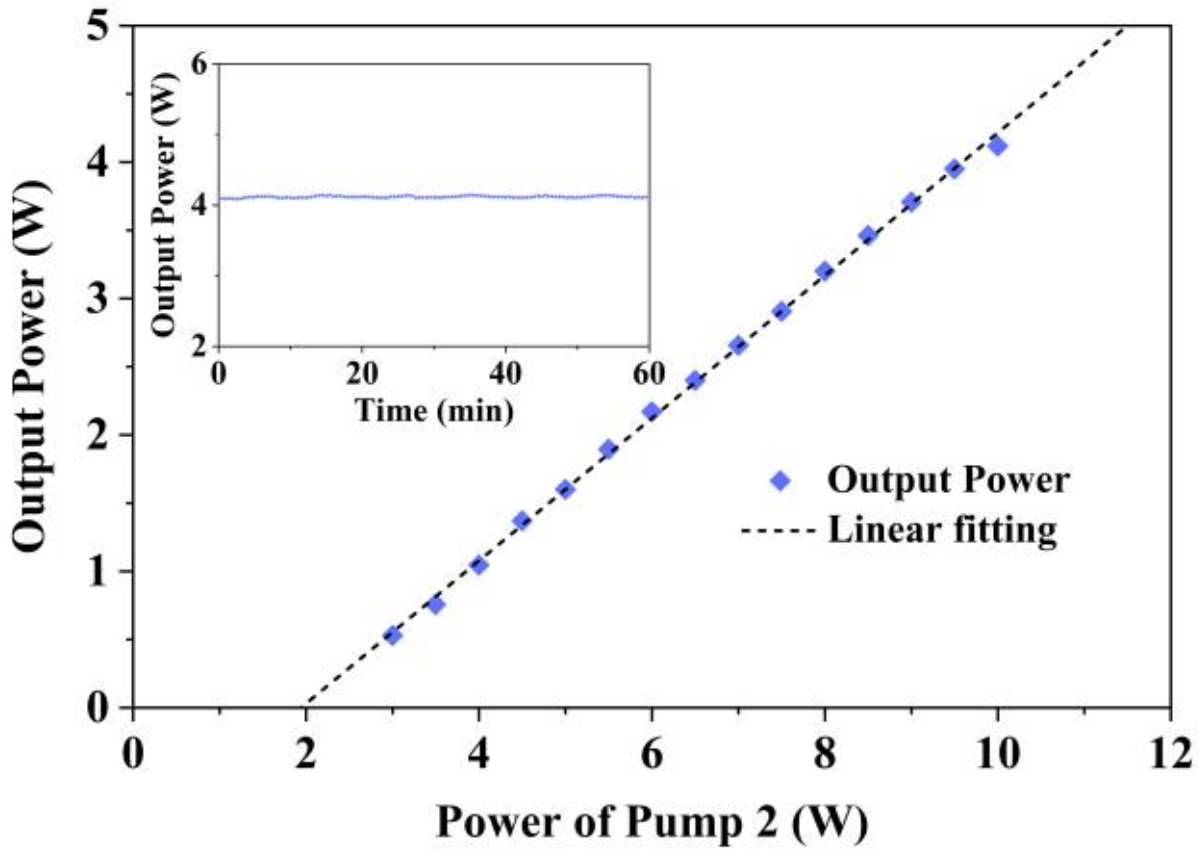

**Figure 3**. Output power with the increasing power of Pump 2. Inset: the output power stability when the power of Pump 2 is 10 W. The power of Pump 1 was kept constant at 1.9 W in all cases.

The optical spectra of laser pulses at maximum output power were measured. As shown in Figure 4(a), the 10-dB bandwidth of output spectrum from PBS 1 was measured to be over 100 nm, indicating Arm 1 operated in a typical GMN regime. Raman scattering is also observed as a low-intensity spectral peak centered at 1115 nm. On other hand, as shown in Figure 4(b), the output spectrum from PBS 3 was centered at 1030.4 nm, and its 10-dB bandwidth was only 7.7 nm. Both its central wavelength and bandwidth are similar to those of CVBG, indicating that the spectrum remains essentially unchanged during the CPA process in Arm 2. Small spikes in both spectra were mainly attributed to the non-ideal transmission spectra of fiber elements used in the cavity. The autocorrelation trace of laser pulses from PBS 3 at maximum output power is shown in Figure 5. Assuming a Gaussian shape, its FWHM was measured to be 1045.1 fs, and hence the pulse width was calculated to be 739.0 fs, consistent with the typical value for a CPA. Moreover, the peak power of laser pulses was calculated to be 0.48 MW.

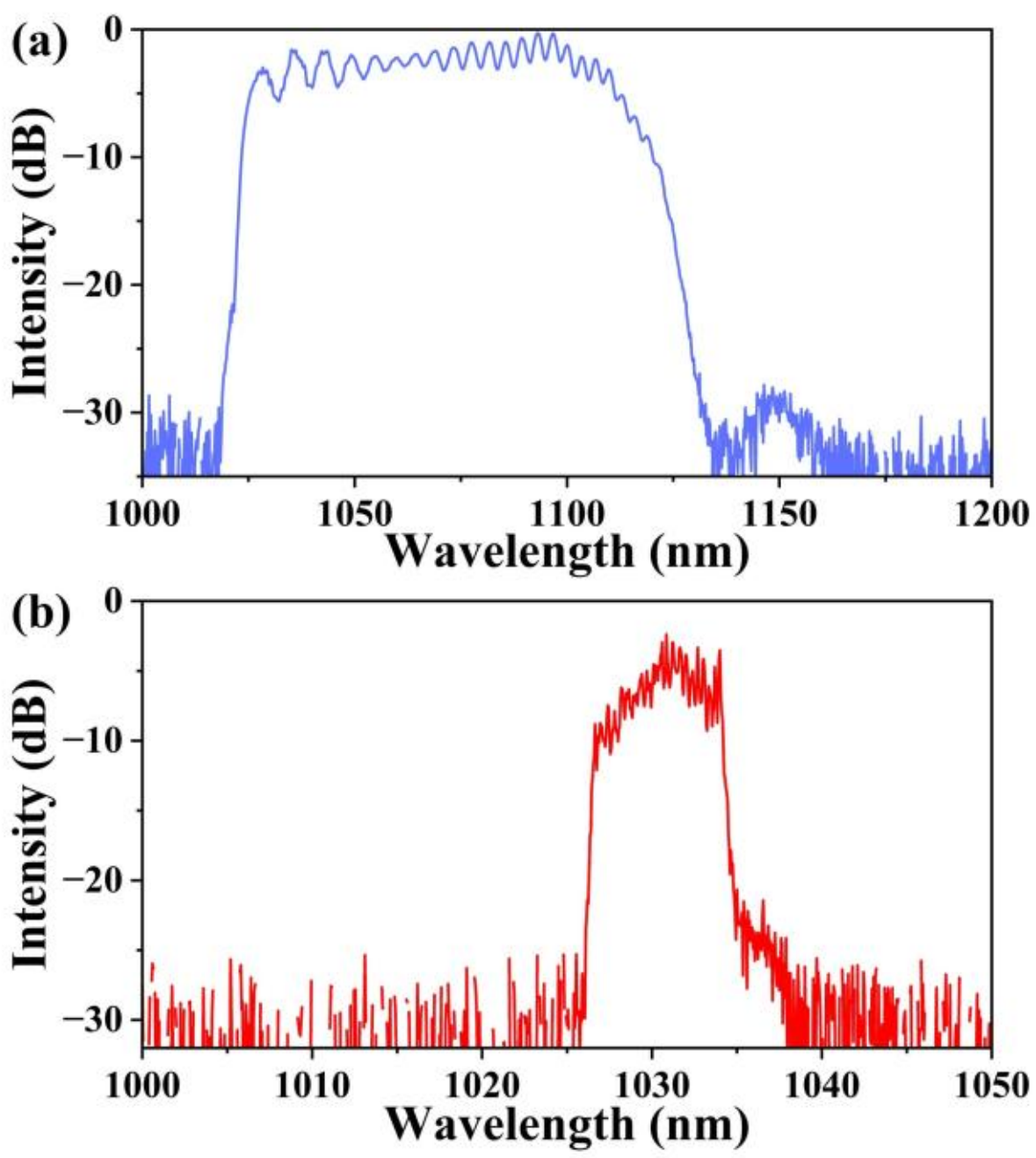

**Figure 4**. (a) The output optical spectrum from PBS 1 with a spectral resolution of 0.2 nm and a span range of 200 nm, and (b) the output optical spectrum from PBS 3 with a spectral resolution of 0.05 nm and a span range of 50 nm.

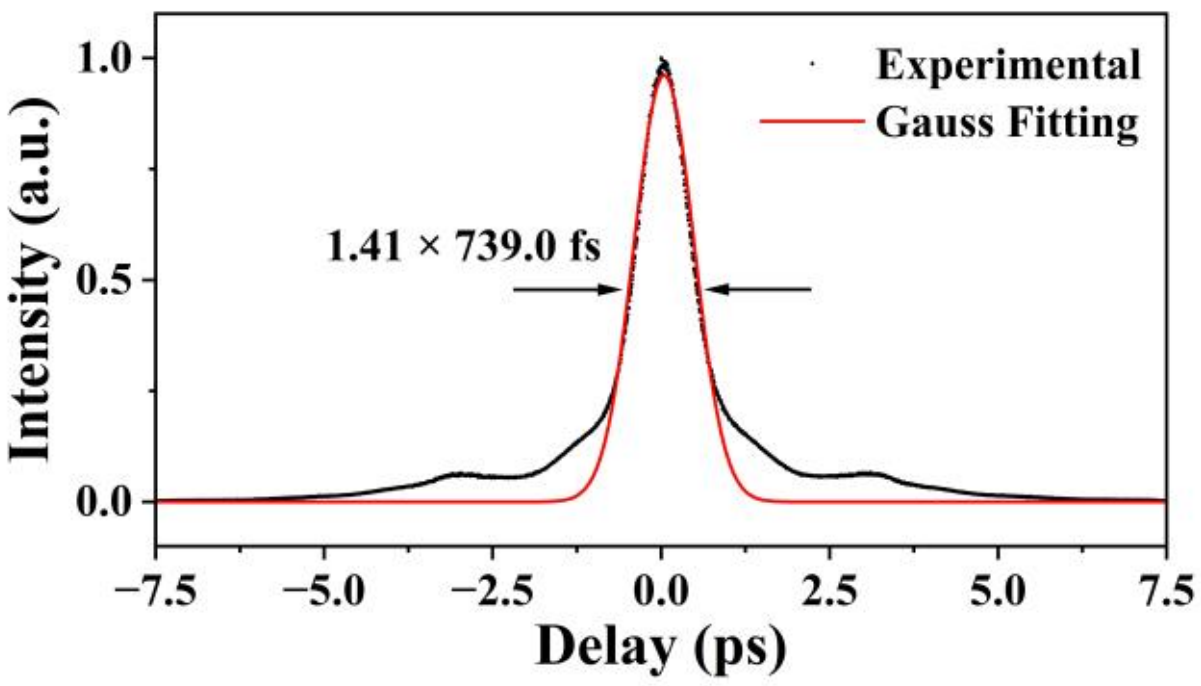

**Figure 5**. The autocorrelation trace of output pulses with Gaussian fitting.

Figure 6 illustrates the radio frequency (RF) spectrum of laser pulses from PBS 3 at the maximum output power. The RF spectral center was measured to be 12.60 MHz, which corresponds to the total cavity length of 23.8 m. The signal-to-noise ratio of the RF spectrum was as high as 65.5 dB; no signal of multi-pulsing or harmonic mode-locking could be found in the long-span RF spectrum. These results further confirm stable fundamental-mode-locking operation of the laser at the maximum output power.

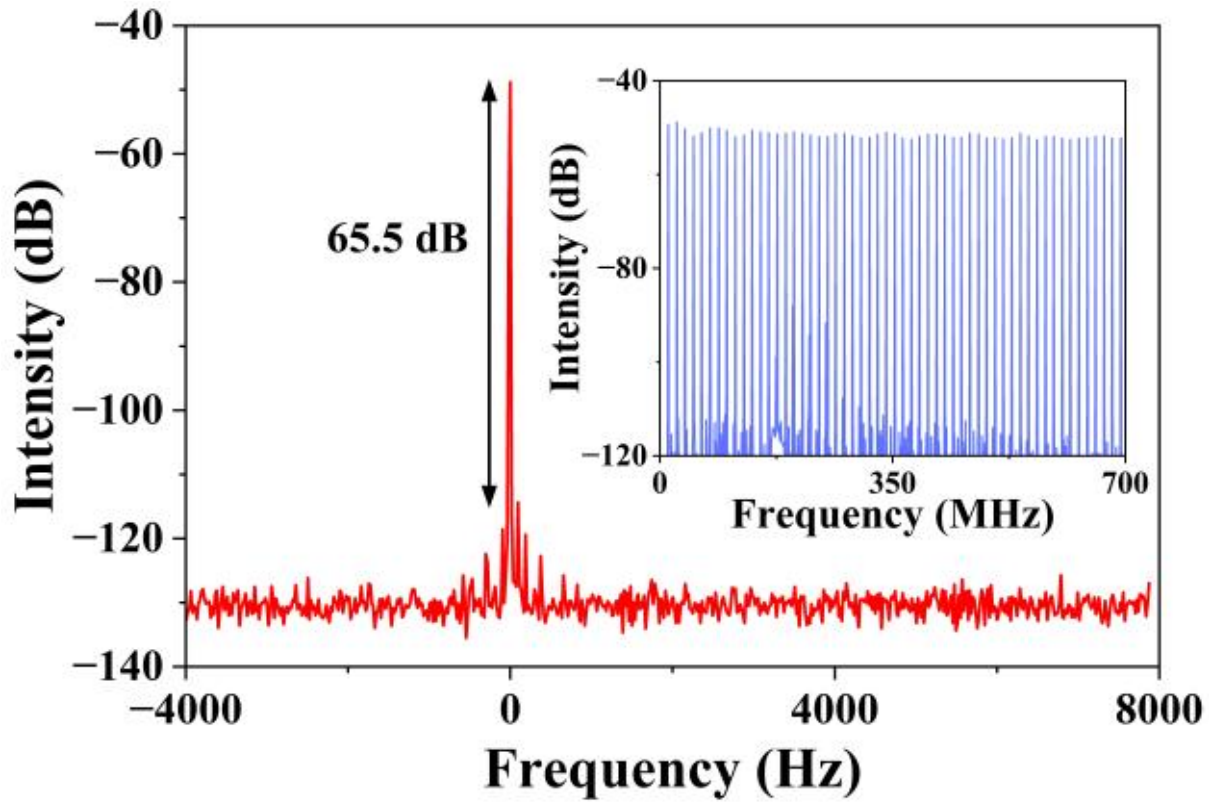

**Figure 6**. The output radio frequency spectrum with a resolution bandwidth of 1 Hz and a span range of 12 kHz; the central frequency is shifted to zero for clarity. Inset: the output radio frequency spectrum with a resolution bandwidth of 100 Hz and a span range of 700 MHz.

# IV. Discussion

Different from those reported MOs, the GMN evolution serves as a nonlinear attractor, and hence a seed provider instead of a power amplifier in our MO. This provides a new approach to constructing the MO system. For example, in addition to the linear CPA evolution shown above, other pulse evolutions such as dispersion-managed soliton, similariton and even mode-locked Raman lasing can be achieved in Arm 2 if the configuration is suitable for these evolutions.

The CPA configuration was chosen in Arm 2 since higher pulse energy can be achieved by reducing the laser peak intensity and hence large nonlinearities. Therefore, the output pulse energy of this MO is much higher than typical MOs using gain fibers with similar mode area [8, 10, 14, 20]. Moreover, it can be scaled further to reach the level of a typical CPA system by using gain fibers with larger mode area and other pre-chirp management.

# V. Conclusion

In conclusion, we have provided a new approach to design a high-energy MO. This MO was based GMN regime in one arm and CPA in the other arm, which is first demonstrated. It can produce ultrashort laser pulses with the pulse energy over 300 nJ and pulse width of 739 fs. Future works will focus on the optimization and further development of this novel design.

# Acknowledgement

Authors thank the EMGO-TECH Ltd. for supporting the fiber components for this work. This work was partially supported by National Natural Science Foundation of China (62205056, U25B2082); The Guangdong Basic and Applied Basic Research Foundation (2024B1515130001).